\documentstyle[12pt]{article} 
\normalbaselines

\newcommand{\be}{\begin{equation}}
\newcommand{\ee}{\end{equation}}
\newcommand{\bea}{\begin{eqnarray}}
\newcommand{\eea}{\end{eqnarray}}
\newcommand{\r}{\rho}
\newcommand{\th}{\varphi}

\def\lb#1{\label{eq:#1}}
\def\rf#1{(\ref{eq:#1})}

\newcommand{\E}{{\cal E}_0}

\begin{document}
\bibliographystyle{unsrt}

\vbox {\vspace{6mm}}

\begin{center}
{\large \bf EXACT, ${\bf E=0}$, CLASSICAL AND QUANTUM SOLUTIONS \\[2mm]
FOR GENERAL POWER-LAW  OSCILLATORS }\\[7mm]
Michael Martin Nieto\footnote{Email:  mmn@pion.lanl.gov}\\
{\it Theoretical Division, Los Alamos National Laboratory\\
University of California\\
Los Alamos, New Mexico 87545, U.S.A. }\\[5mm]

Jamil Daboul\footnote{Email:  daboul@bguvms.bgu.ac.il}\\
{\it Physics Department, Ben Gurion University of the Negev\\
Beer Sheva, Israel}\\[5mm]
\end{center}

\vspace{2mm}

\begin{abstract}

For zero energy, $E=0$,
we derive exact, classical and quantum solutions for {\em all}
power-law
oscillators  with potentials $V(r)=-\gamma/r^\nu$,
 $\gamma>0$ and $-\infty <\nu<\infty$.
When the angular momentum is non-zero,
these solutions lead to the classical orbits  $\r(t)=
 [\cos \mu (\th(t)-\th_0(t))]^{1/\mu}$,
with  $\mu=\nu/2-1 \ne 0$.
For $\nu>2$, the orbits are bound and go through the origin.
We calculate the periods and precessions of these bound orbits, and
graph a number of specific examples.
  The unbound orbits are also discussed in detail.
Quantum mechanically, this system is also  exactly solvable.
We find that
when $\nu>2$ the
solutions are normalizable (bound),
as in the classical case.  Further, there are normalizable discrete,
yet {\it unbound},
states.  They   correspond to unbound classical particles which reach
infinity in a finite time.
Finally, the number of space dimensions of the
system can determine whether or not
an $E=0$ state is bound.
These and other interesting comparisons to
the classical system will be discussed.

\end{abstract}


\section{Introduction}

This all really started in Moscow, in 1992.  That is where I met my colleague,
Jamil Daboul, at the Second International Workshop on Squeezed States
and Uncertainty Relations.  It was held at the Conference Center-Hotel
that the Russian Academy of Sciences uses.  Late at night
Jamil and I would get into  sessions on life, physics, women, politics --
you know, the usual stuff -- while we drank his scotch.

The physics came around to musings about why certain problems can be solved
exactly while others cannot, and the symmetries associated with such problems.
There is a ``folk-theorem" I often think of, which certainly is not exact
but also certainly is intriguing.   This theorem declares that if you
can solve (or not solve) something classically the same is true quantum
mechanically, and {\it visa versa}.

Things stood there until Jamil visited me last year, and we took things up
again.  While wondering about the role of the Runge-Lenz vector in
potential systems, a number of small observations started us down the line:
a) the classical orbit for the attractive $- \gamma/r^4$
 potential with
centripetal potential barrier can be solved exactly;
b)  this  type of quantum system is usually only discussed for
$E \neq 0$;
c) for $E=0$, both the classical and quantum equations are simpler.
Eventually we found out that {\it all} potentials of the form:
\be
V(r) = -~\frac{\gamma}{r^{\nu}} = -\frac{\gamma}{r^{2\mu+2}}~,
{}~~~~~~\gamma > 0~,
 ~~~~~~  - \infty <\nu<\infty~, \lb{pot}
\ee
can be solved exactly,  both classically and quantum mechanically,
for zero binding energy, $E=0$.  --
In what follows,
it will be useful to switch back and forth between the variables
$\nu$ and $\mu$ related by
\be
\mu = (\nu-2)/2, ~~~~~~~ \nu = 2(\mu+1)~.
\ee

Therefore, contrary to the usual scenario of solving a particular potential
for all energies, one could solve an infinite system of potentials for a
particular energy.  The physics that came out was most amusing.
In this piece I will report on this work.   For further details you can
consult a letter on the new results for the quantum system
\cite{dnpla},
as well as longer articles on the classical and quantum physics involved
\cite{dnajpc,dnajpq}.

In Section 2 I will demonstrate  the solution to the classical problem.
Section 3 contains amusing specific examples of some of the classical
trajectories.  The quantum solution will be given in Section 4.  Section 5
contains interesting aspects of the quantum solutions,
and then I give a brief closing comment.

Before continuing, I wish to further parametrize the power-law
potentials as
\be
  V(r) \equiv - \frac{\gamma}{r^\nu}
        \equiv - \E  \frac{g^2}{\r^\nu}
= - \frac{L^2_0}{2 m a^2} \frac{g^2}{\r^\nu}~, ~~~~
\r \equiv r/a~.       \lb{potN}
\ee
The dimensional coupling constant, $\gamma$, is more useful
in classical physics.
The dimensionless coupling constant, $g^2$,
is more useful in quantum physics.
Note, in particular, that the constant $L_0$ becomes $\hbar$ in quantum
physics.
Finally, the ``effective potential," including the angular-momentum
barrier, is
\be
 U(L,r) = \frac{L^2}{2mr^2}  - \frac{\gamma}{r^\nu}~.   \lb{effpot1}
\ee


\section{Classical Solution}

Let us now obtain the classical solution.
By substituting the angular-momentum conservation condition
\be
\dot \th = L/(mr^2)    \lb{amc}
\ee
into the energy conservation condition
\be
E-V= \frac{m}{2} \dot \th^2
      \left[\left(\frac{dr}{d\th }\right)^2 +r^2\right] ~,
      \lb{ec}
\ee
one obtains
\be
\left(\frac{dr}{d\th }\right)^2 +r^2 = \frac{2m(E-V) r^4 }{L^2}~.
\lb{Ncons}
\ee

This is essentially a first-order differential equation, which could
be formally integrated.
However, for  $E=0$, it is much more
efficient to solve Eq. \rf{Ncons} directly.
Converting to the dimensionless variable $\r=r/a$ and substituting
$V$  into Eq. \rf{Ncons}, we obtain
 \be
\left(\frac{d\r}{d\th }\right)^2 +\r^2 = \r^{(4-\nu)}= \r^{(2-2\mu)} .
\lb{Nreq}
\ee
For $\nu = 4$ the right-hand side of this equation is unity, so the
solution is a cosine. This is the circular orbit $\r= \cos \th$
which we will discuss  in the next section.   Guided by this
  we multiply
Eq. \rf{Nreq} by $\r^{2\mu-2}$ to yield
\be
\left(\r^{\mu-1}\frac{d\r}{d\th}\right)^2 +\r^{2\mu}=
\left(\frac{d\r^\mu}{\mu d\th}\right)^2 +\left(\r^\mu\right)^2 = 1 ~.
\lb{dift}
\ee
Now, $\r^\mu$ satisfies the differential equation for
the trigonometric functions. Therefore, the {\em general} solution of Eq.
\rf{dift}
is  given by
\be
 \r^\mu = \cos \; \mu(\th-\th_0) = \cos\left[\frac{\nu -2}{2} (\th
-\th_0)\right]~,
\lb{sol}
\ee
or
\be
\r = \left[\cos\ \mu\left(\th-\th_0 \right)\right]^{1/\mu}=
 \left[\cos\left(\frac{\nu -2}{2} (\th-\th_0)\right)
      \right]^{\frac{2}{\nu-2}}~ .  \lb{solr}
\ee
The phase, $\th_0$, is the integration constant.

Actually, for bound trajectories, which are the case for $\nu > 2$, the
angle  $\th$    and the phase $\th_0$ both change value at the origin.
There a particle is both at the end of a
particular orbit (which starts and ends at the origin)
and also at the beginning of the next orbit.
$\th$ changes value because of the use of polar coordinates and $\th_0$
  does  because of the singular nature of the potentials.
One has to be careful in
matching solutions for bound orbits,
and I refer you to Ref. \cite{dnajpc} for the details.
For now  just note that this problem can be taken care of, and we set
$\th_0=0$ for the first orbit.


\section{Classical Trajectories}

\subsection{Bound trajectories: $2 < \nu$ or $1 < \mu$}

\noindent For $2 < \nu$ or $1 < \mu$, the trajectories go out of
and back in to
the origin in a finite amount of time.  The reason for this is that the
dynamic potential dominates at the origin, but the centripetal barrier
dominates at a finite distance.  The effective potential then asymptotes
to zero from above as $r \rightarrow \infty$.  This is shown in Figure 1.

These bound
orbits have an opening angle at the origin of
\be
\Phi_{\nu} = \frac{2\pi}{\nu-2}=\frac{\pi}{\mu}~.
\ee
\newpage

{}~~
\vspace{3.7in}

\begin{quotation}
FIG. 1.  The effective potential obtained from Eq. \rf{effpot1}
for $\nu = 4$ in units of ${\cal E}_0/2$, as a function of $\r = r/a$.
The form is $U(\r)=4/\r^2 - 1/\r^4$.
\end{quotation}


\noindent The precession per orbit is
\be
P_{\nu}=  \left(\frac{1}{\mu}-1\right)\pi=
\left(\frac{4-\nu}{\nu-2}\right)\pi~,   \lb{P}
\ee
which means that if $\nu$ is a rational fraction, the
trajectory will
close after a finite number of orbits.
The classical period of an orbit is
\be
\tau_{\nu} = \left[\frac{ma^2}{L}\right]\frac{\sqrt{\pi}}{|\mu|}
        \frac{\Gamma(b)}{\Gamma(b + 1/2)} ~, ~~~~~  b \equiv
\frac{1}{\mu}+\frac{1}{2} = \frac{\nu + 2}{2\nu - 4}>0~.
\ee
(Once again, see Ref. \cite{dnajpc} for details.)

Starting with very large $\nu$, the
first orbit describes a very thin petal. The second orbit precesses by
almost $-\pi$, being a thin petal almost on the opposite side of the first
orbit.  As $\nu$ gets smaller, the petals become larger and the precession
per orbit becomes smaller.

For example, the $\nu=8$ case, has three petals.  Here a petal is
$\pi/3$ wide
and the precession per orbit is $-2\pi/3$.  Thus, there are three orbits
before the trajectory closes. Note that here the three petals in a closed
trajectory  cover only half of the opening angle from the origin.
We show this in Figure 2.

The case $\nu=6$ is very interesting.
 The width of a petal is $\pi/2$ and
the precession is $-\pi/2$ per orbit. Here, the width of a petal
and the
precession are exactly such that there is
no overlap and also no ``empty angles."
It  takes four orbits to
close a trajectory.  This is shown in Figure 3. We see that the physical
solution consists of two perpendicular lemniscates (figure-eight
curves composed of two opposite petals).

\newpage


{}~~
\vspace{3.7in}

\begin{quotation}
FIG. 2.    The first three orbits for $\nu$ = 8.  Each orbit is precessed
$-2\pi/3$ from the previous one, so that by the end of the $3$rd orbit,
the trajectory closes.  In this, and later orbits, we show cartesian
coordinates for orientation.

\end{quotation}

\vspace{3.7in}

\begin{quotation}
FIG. 3.  The first four orbits for $\nu$ = 6.  Each orbit is precessed
$-\pi/2$ from the previous one, so that by the end of the $4$th orbit, the
trajectory closes.
\end{quotation}

\newpage

{}~
\vspace{3.7in}

\begin{quotation}
FIG. 4.  The orbit for $\nu=4$.  It is a circle, and repeats itself
continually.
\end{quotation}

\vspace{3.7in}

\begin{quotation}
FIG 5.  The first two orbits for $\nu$ = 3.  Each orbit is precessed
$\pi$ from the previous one, so that by the end of the $2$nd orbit, the
trajectory closes.
\end{quotation}

\newpage


When we reach
 $\nu=4$, the petals have widened so much that they form a circle.
The  circle
 starts at the origin, travels symmetrically about the positive $x$-axis,
and returns to the origin.  The precession is zero,  so the orbit
continually repeats itself.  In Figure 4 we show this orbit.

As $\nu$ becomes less than $4$, we can think of a petal obtaining a width
greater than $\pi$, i.e.,  an orbit consists of
two spirals, one out and one in,  at opposite ends of
the orbit.

Consider the special case $\nu=3$.  The width of the double-spiral  orbit
is still given by the formula for
$\Phi_{\nu}$, and is $2\pi$.  Therefore, the first orbit
begins and ends towards the negative $x$-axis.  The precession is $\pi$,
so the trajectory closes after two orbits.  We show this case in Figure 5.

As $\nu$ approaches $2$.
the spirals become tighter and tighter and the precession (now clockwise)
becomes larger.
In fact, the spirals' angular variation as well as the orbit's precession
both become infinite in magnitude as $\nu$ approaches $2$.

\subsection{Unbound trajectories: $\nu \leq 2$ or $\mu\leq 0$}

When $\nu$ reaches $2$, there is a singular change.  First, the
double spiral becomes infinite in angular width. But also, the joining of
the two sides of the
double spiral at $\r=1$ and $\th=0$ breaks down.   It is as if a
tightly-wound double spring broke.  The ends
spiral out to infinity.  This special case is a Cotes' (infinite) spiral.
It takes an infinite time to reach infinity from the origin.

When the potential
 parameter $\nu$ just leaves
that of the infinite spiral, that is, when
one barely has $\nu < 2$ or $\mu < 0$, there is another change.  Although
the two ends of the entire trajectory still reach to infinity and the
spirals in and out almost have  infinite angular widths, the distance of
closest approach jumps from $\r = 0$ to $\r = 1$.

As  the value of $\nu$ decreases, the value of the angular width
of the trajectory, now
given by  $\Phi_{\nu} = \pi/|\mu|$, also  decreases accordingly.
By the time $\nu = 1$,
the angular width has decreased to $2\pi$.    Eventually it
becomes less than $\pi$, meaning the orbit comes in and
out in the same half plane.  This happens for $\nu < 0$, i.e., when
the force becomes repulsive.

When  $0 < \nu < 2$, the repulsive centripetal barrier dominates at small
$r$ whereas the attractive potential  $V=-\gamma/r^\nu$
dominates at large
$r$. A typical shape  is familiar from the Kepler problem.
Therefore, for $0 < \nu < 2$, the $E=0$ classical orbits are all
unbounded.  The distance, $a$,  now has a completely
different interpretation.  It is now  the
distance of closest approach.  Even so, the formal solution \rf{sol}
remains valid for negative values of $\mu$.

As a first example consider the case $\nu = 3/2$ or $\mu = -1/4$.
This orbit has a total angular width of $4\pi$.  It is shown in the
two drawings of Figure 6.  The large-scale first drawing shows the
trajectory coming in from the top, performing some gyration, and going
out at the bottom.  The small-scale second drawing shows the trajectory
winding around twice near the origin, with the distance of closest
approach being one.

A second example is the  exact Kepler potential,
$\nu=1$ or $\mu=-1/2$.
Eq.  \rf{sol} gives
\be
\r^{-1/2}=\cos \th /2 ~,
\ee
so that
\be
\frac{1}{\r}=(\cos \th /2)^2=\frac{ 1+ \cos \th}{2} ~.
\ee
This is the famous  parabolic orbit for the Kepler
problem with $E=0$.   This orbit is shown in the first drawing
of Figure 7.  The parabola yields an angular width of $2\pi$, as it
should.

\newpage


{}~
\vspace{3.7in}

\begin{quotation}
FIG. 6.   A large-scale view, and a small-scale view near the origin,
of the trajectory for $\nu = 3/2$.
\end{quotation}

\vspace{3.7in}

\begin{quotation}
FIG 7.
  From left to right,
he trajectories for the cases i) $\nu = 1$, ii) $\nu = 0$,
iii) $\nu = -2$, and iv) $\nu = -4$.  The curves are labeled by the
numbers $\nu$.
\end{quotation}

\newpage


If we formally set $\nu=0$ in the expression \rf{potN}, we get a
negative constant potential $V(r)=-\gamma$. Therefore,
in this case the force vanishes
 and we have a free particle.  Its  orbit must be a
straight
 line.  However, Eq. \rf{Nreq} shows that one still has
the same type of  solution, Eq. \rf{solr}.  Here
it is
\be
\r = [\cos\th]^{-1}~, ~~~~~ x = r\cos\th=a ~.
\ee
This is the equation for a vertical straight line that crosses the
$x$-axis at  $x=a$, as required by the initial conditions.
This orbit is shown in the second drawing Figure 7, it subtending an
angular width of $\pi$ from the origin.

For $\nu <0$ or $\mu < -1$
the potentials $V(r)$ in Eq. \rf{potN} are repulsive and
negative-valued for all $r>0$,  with $V(r)$ going to $-\infty$ at large
distances.   Since both the potential, $V(r)$, and the centripetal potential
decrease monotonically, the effective potential has no minima or maxima.
Even so,  for $E=0$ these unbounded orbits  behave qualitatively like those
for $ 0 \leq \nu < 2$. The quantity $a$ now labels the distance of
closest approach  and the solutions are given by the same expression
\rf{sol}, which is also valid for all $\mu < 0$:
\be
\r=[\cos\mu \th]^{1/\mu}= [\cos |\mu| \th]^{-1/|\mu|}~, ~~~~~~\mu < 0~.
\ee

The most famous  special case of these
potentials is the ``inverted"
harmonic-oscillator potential, with $\nu=\mu=-2$.
The orbit  is given by $\r= [\cos 2\th]^{-1/2}~$, so that
\be
1 = \frac {r^2}{a^2} \cos 2\th=\frac {r^2}{a^2}
(\cos^2 \th - \sin^2 \th )= \frac {x^2}{a^2}- \frac {y^2}{a^2} ~.
\ee
Thus, the trajectory
is a  special hyperbolic orbit, whose minor
and major axes are equal, $b^2=a^2$.
We show this orbit as the third drawing in Figure 7.  Now the
angular width has decreased to $\pi/2$.

As the last case, we
consider  the orbit for $\nu = -4$ or $\mu = -3$.  This orbit is shown
in the last drawing of Figure 7.  The orbit subtends an angle of
$\pi/3$, again as it should.  One sees that as $\nu$ becomes more and
more negative, the orbits will become narrower and narrower.  This is
just as in the bound case, where the petals became narrower and
narrower as $\nu$ became more and more positive.


\section{Quantum Solution}

Consider the radial Schr\"odinger equation with angular-momentum
quantum number $l$:
\be
E R_l =\left[-\frac{\hbar^2}{2m}\left(\frac{d^2}{dr^2}
+ \frac{2}{r}\frac{d}{dr}-\frac{l(l+1)}{r^2}\right)
+ V(r)\right] R_l ~.  \label{scheq}
\ee
This Schr\"odinger equation is exactly solvable
for the potential of Eq. \rf{potN}
for all  $E=0$ and all $-\infty <\nu < \infty$.
 To see this, set $E =0$ in Eq. (\ref{scheq}),
change variables to $\r$,
and then multiply by $-\r^2$.  One finds
\be
0=\left[\r^2\frac{d^2}{d\r^2} + 2\r\frac{d}{d\r} - l(l+1)
         + \frac{g^2}{\r^{2\mu}}\right]R_l(r) ~.
\label{qmsp}
\ee

This is a  well-known differential equation of
mathematical physics.  For
$\nu \neq 2$ or $\mu \neq 0$, the solution can be directly
given as
\be
R_l(r) = \frac{1}{\r^{1/2}}
           {\mbox{\huge{J}}}
           _{\left(\frac{2l+1}{|\nu - 2|}\right)}
           \left(\frac{2g}
           {|\nu - 2|\r^{\left(\frac{\nu - 2}{2}\right)}}
           \right)
     = \frac{1}{\r^{1/2}}
           {\mbox{\huge{J}}}
           _{\left(\frac{l+1/2}{|\mu|}\right)}
           \left(\frac{g}
           {|\mu| \r^{\mu}}\right)~, ~~~ \mu \neq 0~.  \label{gensol}
\ee
One actually has to be careful about when an absolute value of $\mu$ is
called for in the labels of the solution, and whether the $J$ Bessel
functions are called for {\it vs}. the $Y$ function.  These details are given
in Ref. \cite{dnajpq}.


\section{Properties of the Quantum Solution}

\subsection{Normalizable bound states:  $2 < \nu$ or $1 < \mu$}

The normalization constants for the wave
functions would have to be of the form
\be
N_l^{-2} = \int_{0}^{\infty}\frac{r^2dr}{\r}
             {\mbox{\huge{J}}}^2
           _{\left(\frac{l+1/2}{|\mu|}\right)}
           \left(\frac{g}
           {|\mu| \r^{\mu}}\right)~.\label{norm}
\ee
Changing variables first from $r$ to $\r$ and then from $\r$ to
$z=g/(|\mu|\r^{\mu})$, and being
careful about the limits of integration for all $\mu$, one obtains
\be
N_l^{-2} = \frac{a^3}{|\mu|}\left(\frac{g}{|\mu|}\right)^{2/\mu}~
I_l~~,
\ee
where
\be
I_l = \int_0^{\infty}\frac{dz}{z^{(1+2/\mu)}}
              {\mbox{\huge{J}}}^2
           _{\left(\frac{l+1/2}{|\mu|}\right)}
           (z) ~. \lb{int}
\ee
This   integral
 is convergent and given by
\be
I_l= \frac{1}{2\pi^{1/2}}
\frac{\Gamma\left(\frac{1}{2}+\frac{1}{\mu}\right)}
{\Gamma\left(1+\frac{1}{\mu}\right)}
\frac{\Gamma\left(\frac{l+1/2}{|\mu|}-\frac{1}{\mu}\right)}
{\Gamma\left(1+\frac{l+1/2}{|\mu|}+\frac{1}{\mu}\right)}~,
\lb{norm}
\ee
if the following  two conditions are satisfied:
\be
\frac{2l+1}{|\mu|} + 1> \frac{2}{\mu}+ 1>0~.  \lb{gt}
\ee

Eqs. \rf{norm} and \rf{gt} lead to two sets of
normalizable states.   The first is when
\be
\mu > 0 ~~~ \mbox{or} ~~~\nu > 2~,   ~~~~~~~  l >1/2~.
\lb{lll}
\ee
  These are ordinary
 bound states and result because the effective
 potential asymptotes to zero from above, as in Figure 1.
In this case, for $E=0$, the wave function can reach infinity
only by tunneling through an infinite forbidden region.
 That takes forever, and so the state is bound.

Note that the condition on $l$
in Eq. \rf{lll} is the minimum nonzero angular momentum
allowed in quantum mechanics, $l_{min} = 1$.  This agrees with the
classical orbit solution which is bound for any nonzero
angular momentum.
Also, the above $E=0$ solutions exist for all $g^2 > 0$, and not
just for discrete values of the coupling constant.

\subsection{Free states:  $-2 \leq \nu \leq 2$ or $-2 \leq \mu \leq 0$}

For $-2 \leq \nu \leq 2$ or $-2 \leq \mu \leq 0$ and $l\ge 1$
(as well as the
solutions with $l=0$ and $0 < \mu$ or $2 < \nu $) the solutions are
free, continuum solutions.  This is in analogy to the classical case,
where the trajectories are normal and free.

\subsection{Unbound yet normalizable states:  $\nu < -2$ or $\mu<-2$}

There is another class of
normalizable solutions which is quite surprising.
For any $l$ and all $\nu < -2$ or $\mu<-2$, one can
verify that  the conditions of Eq. \rf{gt} are also satisfied.
Thus, even though one here has a repulsive potential that falls
off faster than the inverse-harmonic oscillator and the
states are {\it not} bound, the solutions are
{\it normalizable}!

The corresponding classical solutions yield
infinite orbits, for which the particle needs only a finite time
to reach infinity \cite{dnajpc}.
But it is known that a classical potential which yields
trajectories with a finite travel time to infinity
also yields a discrete spectrum
in the quantum case.
This discrete spectrum is obtained by imposing particular boundary
conditions on
the solutions, which defines a self-adjoint extension of the
Hamiltonian.  (See Ref. \cite{dnpla}.)

\subsection{Bound states in arbitrary dimensions}

One can easily generalize the problem of the last
section to arbitrary $D$
space  dimensions.   Doing so yields another surprising
physical result.

To obtain the
$D$-dimensional analogue of Eq. (\ref{qmsp}),  one simply has to
replace $2\r$ by $(D-1)\r$ and $l(l+1)$ by $l(l+D-2)$.
The solutions  follow similarly as
\bea
R_{l,D} &=& \frac{1}{\r^{D/2-1}}
           {\mbox{\huge{J}}}
           _{\left(\frac{2l+D-2}{|\nu - 2|}\right)}
           \left(\frac{2g}
           {|\nu - 2|\r^{\left(\frac{\nu - 2}{2}\right)}}
           \right)
                        \\ \nonumber
     &=& \frac{1}{\r^{D/2-1}}
           {\mbox{\huge{J}}}
           _{\left(\frac{l+D/2-1}{|\mu|}\right)}
           \left(\frac{g}
           {|\mu| \r^{\mu}}\right)~.
\eea

To find out which states are now normalizable one first has to change
the integration measure  from $r^2dr$ to $r^{D-1}dr$ and then
continue as before. The end
result is that if the wave functions are normalizable,
the normalization constant is given by
\be
N_{l,D}^{-2} =\frac{a^D}{|\mu|}\left(\frac{g}{|\mu|}\right)^{2/\mu}~
I_{l,D}~~,
\ee
where
\be
I_{l,D} = \frac{1}{2\pi^{1/2}}
\frac{\Gamma\left(\frac{1}{2}+\frac{1}{\mu}\right)}
{\Gamma\left(1+\frac{1}{\mu}\right)}
\frac{\Gamma\left(\frac{l+D/2-1}{|\mu|}-\frac{1}{\mu}\right)}
{\Gamma\left(1+\frac{l+D/2-1}{|\mu|}+\frac{1}{\mu}\right)}~,
\ee
which is defined and convergent for
\be
\frac{2l+D-2}{|\mu|}+1 >\frac{2}{\mu}+1>0~.
 \lb{gtD}
\ee

This yields the surprising result that there are bound states for
all $\nu > 2$ or $\mu > 0$ when $l >2-D/2$.  Explicitly this
means that the minimum allowed $l$ for there to be zero-energy
bound states are:
\bea
D = 2~, ~~~~~ l_{min} &=& 2~, \nonumber \\
D = 3~, ~~~~~ l_{min} &=& 1~, \nonumber \\
D = 4~, ~~~~~ l_{min} &=& 1~, \nonumber \\
D > 4~, ~~~~~ l_{min} &=& 0~. \lb{lmin}
\eea

This effect of dimensions  is purely quantum mechanical, and
exists for all central potentials.
  Classically, the number of dimensions involved
in a central potential problem has no intrinsic effect on the
dynamics.  The orbit remains in two dimensions, and the problem is
decided by the form of the effective potential, U, which contains
only the angular momentum barrier and the dynamical potential.

In quantum mechanics there are two places where
an effect of dimension appears.  The first is in the factor $l(l+D-2)$
of the angular-momentum
barrier.  The second is more
fundamental. It is due to the operator
\be
U_{qm} = -\frac{(D-1)}{\r}\frac{d}{d\r}~.
\ee
This is a new contribution  to the ``effective potential,"
and can be calculated \cite{dnpla}.  The end result is that
given in Eq.  \rf{lmin}.

The dimensional effect  produces what amounts to an additional
centrifugal barrier
which can bind the
wave function at the threshold, even though the
expectation value of the angular momentum vanishes.


\section{Closing Comment}

I hope you have found this discussion of anharmonic power potentials
entertaining and enlightening.  Jamil and I certainly have.
The intuition obtained into the workings and relationships between
classical and quantum physics has been delightful to us, to say the
least.

Thank you very much.



\begin{thebibliography}{99}

\bibitem{dnpla}  J. Daboul and M. M. Nieto, Phys. Lett. A (submitted).

\bibitem{dnajpc}  J. Daboul and M. M. Nieto, Am. J. Phys., in preparation,
on the classical solutions  of the $E=0$, power-law system.

\bibitem{dnajpq}  J. Daboul and M. M. Nieto, Am. J. Phys., in preparation,
on the quantum solutions  of the $E=0$, power-law system.

\end {thebibliography}
\end{document}